\documentclass[12pt]{iopart}
\usepackage{graphicx}
\begin{document}

\title{Self-averaging and criticality: A comparative study in 2d random bond spin models}

\author{N G Fytas and A Malakis}

\address{Department of Physics, Section of Solid State
Physics, University of Athens, Panepistimiopolis, GR 15784
Zografos, Athens,
Greece}\eads{\mailto{nfytas@phys.uoa.gr},\mailto{amalakis@phys.uoa.gr}}

\begin{abstract}
We investigate and contrast, via the Wang-Landau (WL) algorithm,
the effects of quenched bond randomness on the self-averaging
properties of two Ising spin models in 2d. The random bond version
of the superantiferromagnetic (SAF) square model with nearest- and
next-nearest-neighbor competing interactions and the corresponding
version of the simple ferromagnetic Ising model are studied. We
find that, the random bond SAF model shows a strong violation of
self-averaging, much stronger than that observed in the case of
the random bond Ising model. Our analysis of the asymptotic
scaling behavior of the variance of the distribution of the
sample-dependent pseudocritical temperatures is found to be
consistent with the renormalization group prediction of Aharony
and Harris. Using this alternative approach, we find estimates of
the correlation length exponent $\nu$ in agreement with results
obtained from the usual finite-size scaling (FSS) methodology.
\end{abstract}

\vspace{2pc} \noindent{\it Keywords}: Classical Monte Carlo
simulations, Classical phase transitions (Theory), Finite-size
scaling, Disordered systems (Theory) \maketitle

\section{Introduction}
\label{sec:1}

Phase transition in systems with quenched disorder are of
considerable theoretical and experimental interest and have been
the subject of intensive
investigations~\cite{stinchcombe-93,shalaev-94,selke-94}.
Nowadays, the effect of disorder coupled to the energy density on
second-order phase transitions is well understood. The phase
transition remains second-order and the eventual modification of
the universality class is governed by the specific heat divergence
exponent $\alpha_{p}$, as stated by the Harris
criterion~\cite{harris-74}. According to this criterion, if
$a_{p}$ is positive, disorder is relevant, i.e. the system will
reach a new critical behavior under the presence of quenched
randomness. Otherwise, if $a_{p}$ is negative, disorder is
irrelevant and the critical behavior will not change. The value
$\alpha_{p}=0$ of the 2d Ising model is an inconclusive, marginal
case~\cite{MK-99}.

In the present paper we simulate two different 2d random bond
Ising spin systems, the marginal case of the simple 2d Ising model
($\alpha_{p}=0$) and the non-marginal case of the square SAF model
with nearest- ($J_{nn}$) and next-nearest-neighbor ($J_{nnn}$)
competing interactions on the square
lattice~\cite{swendsen-79,binder-80} for $R=J_{nn}/J_{nnn}=1$
($\alpha_{p}>0$~\cite{malakis-06}). Several aspects of the
critical behavior of these two models have been elucidated in
references~\cite{fytas-08b,fytas-08c} and will be also briefly
reviewed in the following Section. Our aim here is to study and
contrast the dependence of the self-averaging properties of the
systems and use these properties for an alternative investigation
of their critical behavior. Self-averaging is an important issue
when studying phase transitions in disordered systems. Although it
has been known for many years now that for (spin and regular)
glasses there is no self-averaging in the ordered
phase~\cite{binder-86}, for random ferromagnets such a behavior
was first observed for the random field Ising model in a paper by
Dayan \etal~\cite{dayan-93} and some years later for the random
versions of the Ising and Ashkin-Teller models by Wiseman and
Domany~\cite{WD-95}. These latter authors proposed a FSS ansatz
describing the absence of self-averaging and the universal
fluctuations of random systems near critical points that was
refined and put on a more rigorous basis by the intriguing
renormalization group work of Aharony and Harris~\cite{AH-96}.
Ever since, the subject of breakdown of self-averaging is an
important aspect in several numerical investigations of disordered
spin
systems~\cite{EB-96,PSZ-97,WD-98,BF-98,TO-01,PS-02,BC-04,MG-05,fytas-06,GL-07}.

In the next Section we define the models under study and we review
their critical behavior. Basic analytical predictions on the issue
of self-averaging in disordered systems are also presented in this
Section. In Section~\ref{sec:3} we briefly outline our entropic
sampling scheme and we present our results on the self-averaging
properties of both models. Finally, we summarize our conclusions
in Section~\ref{sec:4}.

\section{Models and theoretical aspects on self-averaging}
\label{sec:2}

The general form of the Hamiltonian of the random bond versions of
both models considered here is given by
\begin{equation}
\label{eq:1}
\mathcal{H}=\sum_{<i,j>}J_{ij}^{(nn)}S_{i}S_{j}+\sum_{(i,j)}J_{ij}^{(nnn)}S_{i}S_{j},
\end{equation}
where $S_{i}=\pm 1$ are Ising spins, the superscripts (nn) and
(nnn) stand for nearest- and next-nearest-neighbor interactions,
and the implementation of bond disorder follows the binary
distribution
\begin{equation}
\label{eq:2}
P(J_{ij})=\frac{1}{2}[\delta(J_{ij}-J_{1})+\delta(J_{ij}-J_{2})];\;\;
\frac{J_{1}+J_{2}}{2}=1;\;\;r=\frac{J_{2}}{J_{1}},
\end{equation}
where the ratio $r$ is the disorder strength and through out the
paper takes the value $r=3/5=0.6$. From equation~(\ref{eq:1}) we
obtain:

\begin{itemize}

\item{The random bond ferromagnetic Ising model for
$J_{ij}^{(nn)}<0$ and $J_{ij}^{(nnn)}=0$. Note here that, with
this distribution the random Ising system exhibits a unique
advantage that its critical temperature $T_{c}$ is exactly
known~\cite{fisch-78} as a function of the disorder strength $r$
through: $\sinh{(2J_{1}/T_{c})}\sinh{(2rJ_{1}/T_{c})}=1$
($k_{B}=1$). This provides an excellent check for the accuracy of
the numerical scheme, by comparing the estimated critical
temperature with the exact result, as illustrated in
reference~\cite{fytas-08b}. Two main and mutually excluded
scenarios~\cite{MK-99} exist for the description of the critical
behavior of the random version of the Ising model. The first, is
the logarithmic corrections scenario, which, based on quantum
field theory results~\cite{DD-81,Shalaev-84,Shankar-87,Ludwig-87},
states that the presence of quenched disorder changes the critical
properties of the system only through a set of logarithmic
corrections to the pure system behavior. The second scenario of
the so-called weak
universality~\cite{KP-94,Kuhn-94,suzuki-74,gunton-75}, assumes
that critical quantities, such as the zero field susceptibility,
magnetization, and correlation length display power law
singularities, with the corresponding exponents $\gamma$, $\beta$,
and $\nu$ changing continuously with the disorder strength;
however this variation is such that the ratios $\gamma/\nu$ and
$\beta/\nu$ remain constant at the pure system's value. The
specific heat of the disordered system is, in this case, expected
to saturate. Overall, for the random Ising model, the results of
most studies~\cite{wang-90,ball-97,reis-97,selke-98,kenna-08},
including our recent investigation~\cite{fytas-08b}, support the
scenario of logarithmic corrections. Here, additional evidence in
favor of this scenario are presented, via an alternative route
that involves the asymptotic scaling behavior of the
sample-to-sample fluctuations of the susceptibility's
pseudocritical temperature.}

\item{The random bond square ($R=1$) SAF model for
$J_{ij}^{(nn)}=J_{ij}^{(nnn)}>0$. It is well-known that the pure
square SAF model develops at low temperatures SAF order for
$R=J_{nn}/J_{nnn}>0.5$~\cite{swendsen-79,binder-80}. For the case
$R=1$, that we deal with, the pure system undergoes a second-order
phase transition, in accordance with the commonly accepted
scenario of a non-universal critical behavior with exponents
depending on the coupling ratio $R$~\cite{swendsen-79,binder-80}.
In fact, recent numerical studies of the model indicated that the
$R=1$ model undergoes, at its pure
version~\cite{malakis-06,monroe-07}, a clear second-order
transition with an exponent $\nu_{p}$ very close to that of the 2d
three-state Potts model $\nu_{p}$(Potts)$=5/6$~\cite{wu-82}, and
at its random bond version~\cite{fytas-08b,fytas-08c}, a
transition governed by a random fixed point with a correlation
length exponent $\nu=1.080(20)$ and magnetic exponents that
satisfy the weak universality scenario for disordered systems as
stated by Kim~\cite{kim-96}. Furthermore, a strong saturating
behavior of the specific heat was observed that distinguished this
case of competing interactions from other 2d random bond
ferromagnetic systems studied previously~\cite{kim-96,picco-96}.}

\end{itemize}

Our numerical studies of disordered systems are carried out near
their critical points using finite samples; each sample $i$ is a
particular random realization of the quenched disorder. A
measurement of a thermodynamic property $X$ yields a different
value for the exact thermal average $X_{i}$ of every sample $i$.
In an ensemble of disordered samples of linear size $L$ the values
of $X_{i}$ are distributed according to a probability distribution
$P(X)$. In most Monte Carlo studies, see also our recent
investigations~\cite{fytas-08b,fytas-08c}, one considers the
ensemble averages $[X]_{av}$ and their scaling properties.
However, here, we will focus on obtaining direct evidence about
the nature of the governing fixed points and the self-averaging
properties of the disordered systems via the distribution $P(X)$,
where $X$ may be the sample-dependent pseudocritical temperatures
$T_{c}(i,L)$ and the corresponding magnetic susceptibility peaks
$\chi^{max}(i,L)$. The behavior of this distribution is directly
related to the issue of self-averaging mentioned in the
introduction. In particular, by studying the behavior of the width
of $P(X)$ with increasing the system size $L$, one may address
qualitatively the issue of self-averaging, as has already been
stressed by previous authors~\cite{WD-98}. In general, we
characterize the distribution $P(X)$ by its average $[X]_{av}$ and
also by the relative variance $R_{X}=V_{X}/[X]_{av}^{2}$, where
$V_{X}=[X^{2}]_{av}-[X]_{av}^{2}$. Suppose now that $X$ is a
singular density of an extensive thermodynamic property, such as
$M$ or $\chi$, or the singular part of $E$ and $C$. The system is
said to exhibit self-averaging if $R_{X}\rightarrow 0$ as
$L\rightarrow \infty$. If $R_{X}$ tends to a non-zero value, i.e
$R_{X}\rightarrow const \neq 0$ as $L\rightarrow \infty$, then the
system exhibits lack of self-averaging. The importance of the
above concept has been illustrated by Aharony and
Harris~\cite{AH-96} and their main conclusions are summarized
below:

\begin{itemize}

\item{Away from the critical temperature: $R_{X}=0$. In a finite
geometry, the correlation length $\xi$ is finite for $T\neq T_{c}$
and it can be found, using general statistical arguments,
originally introduced by Brout~\cite{brout-59}, that $R_{X}\propto
(\xi/L)^{d}\rightarrow 0$, as $L\rightarrow \infty$. This is
called strong self-averaging.}

\item{At the critical temperature there exist two possible
scenarios: (i) models in which according to the Harris criterion
the disorder is relevant ($\alpha_{p}>0$): $R_{X}\neq 0$. Then,
the system at the critical point is not self-averaging and (ii)
models in which according to the Harris criterion disorder is
irrelevant ($\alpha_{p}<0$): $R_{X}=0$. In this case $R_{X}$
scales as $L^{\alpha/\nu}$, where $\alpha$ and $\nu$ are the
critical exponents of the pure system, which are the same in the
disordered one. This is called weak self-averaging.}

\item{The pseudocritical temperatures $T_{c}(i,L)$ of the
disordered system are distributed with a width $\delta
[T_{c}(L)]_{av}$, whose square scales with the system size as
$\delta^{2} [T_{c}(L)]_{av}\sim L^{-n}$, where $n=d$ or $n=2/\nu$,
depending on whether the disordered system is controlled by the
pure or the random fixed point, respectively. The above behavior
is now well established by the pioneering works of Aharony and
Harris~\cite{AH-96} and Wiseman and Domany~\cite{WD-95,WD-98}.}

\end{itemize}

In the following Section, we will test the above theoretical
predictions for the two disordered models under study and extract
useful information for some aspects of their critical behavior, as
these emerge from the above described theory.

\section{Simulations and results}
\label{sec:3}

\begin{figure}[ht]
\centerline{\includegraphics*[width=12 cm]{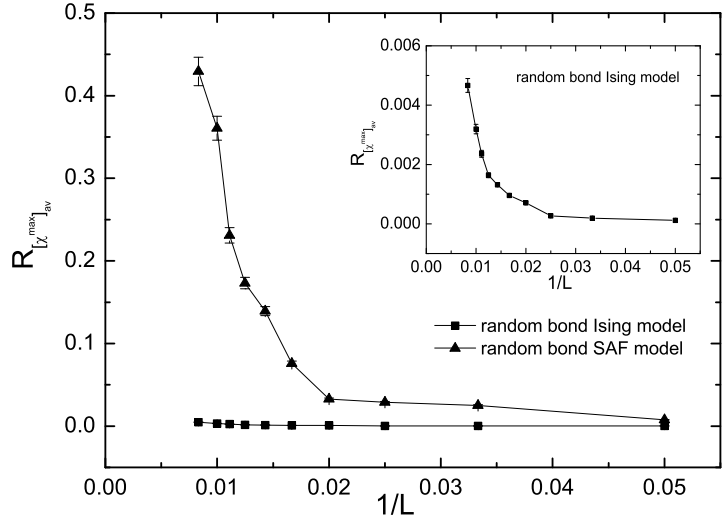}}
\caption{\label{fig:2} Behavior of the ratio
$R_{[\chi^{max}]_{av}}$, defined in the text, as a function of the
inverse lattice size for both models in the main panel and only
for the random bond Ising model in the inset. The statistical
error bars shown are due to the finite number disorder sampling
and have been estimated using the jackknife
method~\cite{newman-99}.}
\end{figure}

\begin{figure}[ht]
\centerline{\includegraphics*[width=16 cm]{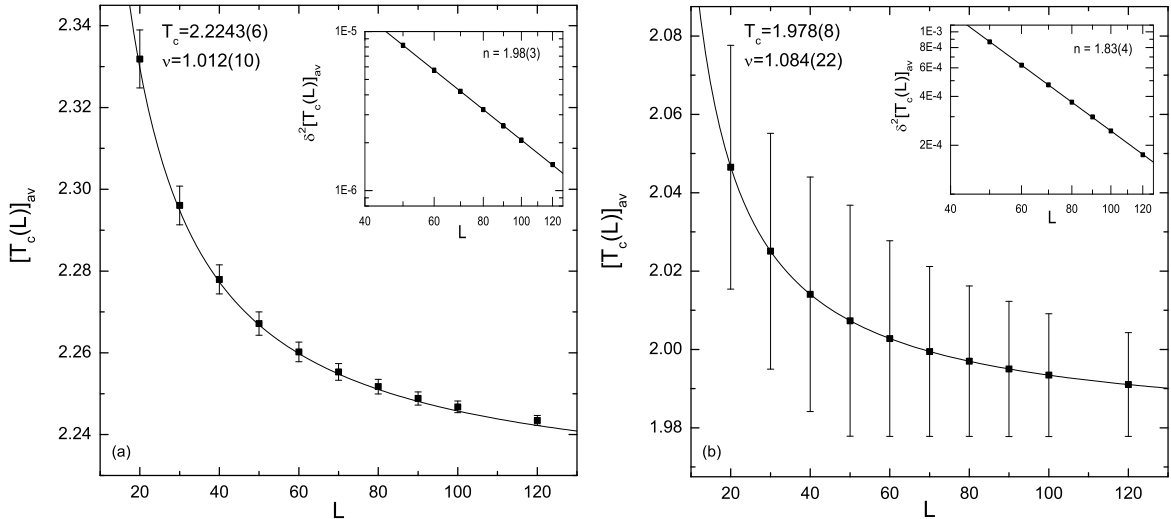}}
\caption{\label{fig:3} Shift behavior of the pseudocritical
temperatures of the magnetic susceptibility for (a) the random
bond Ising model and (b) the random bond square SAF model. The
error bars show the sample-to-sample fluctuations of the models in
the ensembles of the simulated realizations. The corresponding
insets show the FSS of the square of these sample-to-sample
fluctuations in a log-log scale for the larger sizes studied
($L\geq 50$). The solid lines shown are linear fits giving
estimates for the exponent $n$.}
\end{figure}

The numerical scheme used to estimate here the self-averaging
properties of the random versions of the Ising and the SAF models
has been presented in detail in
references~\cite{fytas-08b,fytas-08c}, where a two stage strategy
of an energy-restricted~\cite{malakis-04} implementation of the WL
algorithm~\cite{WL-01} was proposed and successfully applied on
the above random models. In these papers our analysis focused on
the averaged curves $[\ldots]_{av}$ and in particular on the
scaling behavior of their maxima $[\ldots]_{av}^{max}$ , which as
discussed above is the common practise in disordered systems.
Here, we follow a different averaging process, namely that of
averaging over the individual maxima $[\ldots^{max}]_{av}$ of the
random realizations. Within this practise, apart from having an
also accurate, as will be seen below, estimator of mean values,
one has the advantage of estimating directly the corresponding
sample-to-sample fluctuations of a thermodynamic quantity $X$,
that may be defined with the help of the variance of the
corresponding distribution $P(X)$, as defined in the discussion of
self-averaging in Section~\ref{sec:2}. In particular, we consider
square lattices with periodic boundary conditions and linear sizes
$L$ in the range $L=20-120$. For each lattice size we present
results of ensembles of up to $200$ - for the larger sizes - bond
disorder realizations. Note here that, the statistical errors of
the WL method have been omitted from all our figures below, since
they were found to be much smaller than the errors due to the
finite number of disorder averaging, shown in figure~\ref{fig:2},
and even smaller than the sample-to-sample fluctuations, shown in
figure~\ref{fig:3}.
\begin{figure}[ht]
\centerline{\includegraphics*[width=16 cm]{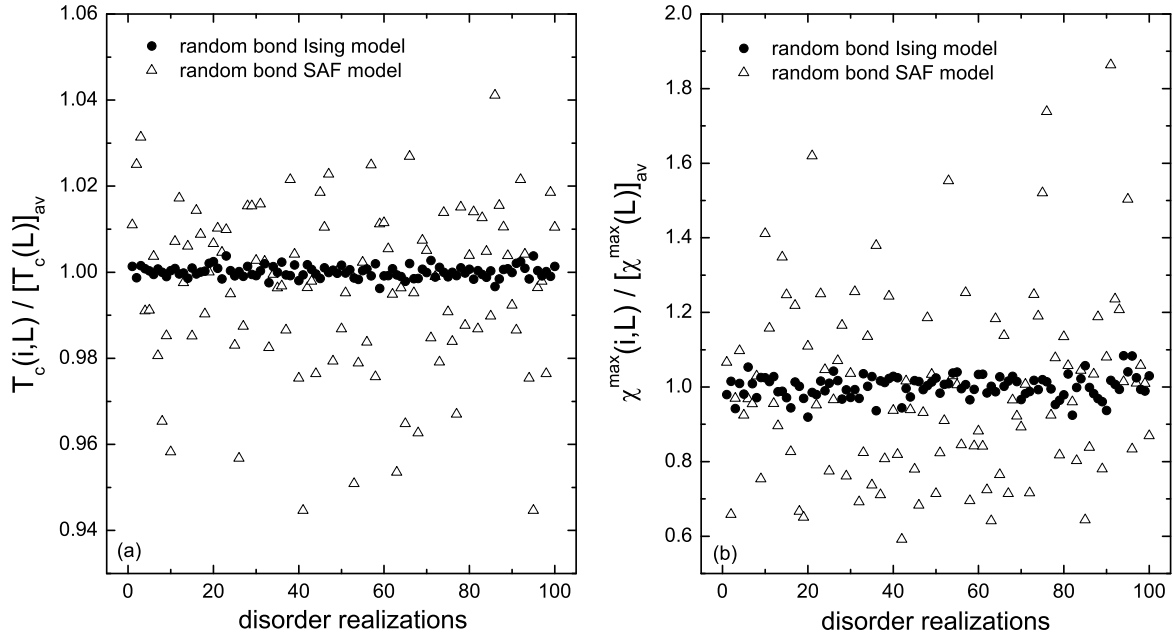}}
\caption{\label{fig:1} Normalized fluctuation of (a) the
pseudocritical temperature and (b) the susceptibility maxima for
both random bond models considered for a lattice size $L=60$ in a
subset of $100$ disorder realizations.}
\end{figure}

We start the presentation of our results with figure~\ref{fig:1},
where we show the normalized fluctuation of (a) the pseudocritical
temperature of the magnetic susceptibility
$T_{c}(i,L)/[T_{c}(L)]_{av}$ and (b) the susceptibility maxima
$\chi^{max}(i,L)/[\chi^{max}(L)]_{av}$ for both random bond models
considered, for a lattice size $L=60$ and a subset of $100$
disorder realizations. It is clear from this figure that for both
quantities, $T_{c}(i,L)$ and $\chi^{max}(i,L)$, the variance of
the data for the case of the random bond SAF model is much larger,
compared to the simple random bond Ising model. This strong
sample-dependence for the random bond SAF model is reflected in
figure~\ref{fig:2}, where we illustrate the FSS behavior of the
ratio $R_{[\chi^{max}]_{av}}$, defined in Section~\ref{sec:2}, as
a function of the inverse linear size for both models considered.
From the main panel and the corresponding inset we observe a
definite approach to a non-zero limiting value, indicating a
strong violation of self-averaging~\cite{AH-96,WD-98,GL-07} of
both the random Ising and SAF models. However, for the case of the
random SAF model this limiting value is much larger, by a factor
of $\sim 100$.
Let us point out here that, for the magnetic susceptibility, the
lack of self-averaging may be present even for pure
systems~\cite{binder-98}. The results shown in the figures above
verify this expectation in a definite way and reveal an
interesting aspect, which is the sensitivity of the microscopic
interactions to bond randomness.

In the main panels of figure~\ref{fig:3} we illustrate the shift
behavior of the individual averaged pseudocritical temperatures
$[T_{c}(L)]_{av}$ of the magnetic susceptibility for (a) the
random bond Ising model and (b) the random bond SAF model. The
error bars shown in this figure, as was already mentioned above,
are the sample-to-sample fluctuations of the two disordered models
$\delta T_{c}(L)$. The shift behavior of these pseudocritical
temperatures is expected to provide estimates for both the
critical temperature and the correlation length exponent, via the
relation
\begin{equation}
\label{eq:3} [T_{c}(L)]_{av}=T_{c}+bL^{-1/\nu}.
\end{equation}
The solid lines shown in both panels are fits of the
form~(\ref{eq:3}) giving the values: $T_{c}=2.2243(6)$ and
$\nu=1.010(12)$ for the random bond Ising model and
$T_{c}=1.978(8)$ and $\nu=1.084(22)$ for the random bond SAF
model. For the case of the random Ising model, the estimated value
$T_{c}=2.2243(6)$ of the critical temperature is in excellent
agreement with the value $2.2245(7)$, obtained in
reference~\cite{fytas-08b} from the scaling of the ensemble
average $[\ldots]_{av}^{max}$, via a simultaneous fitting of the
specific heat's and magnetic susceptibility's pseudocritical
temperatures, and also with the exact value $T_{c}=2.22419\ldots$.
Note that, this latter comparison consists a concrete reliability
test in favor of the accuracy of our numerical data. Also, for the
correlation length exponent, our results indicate that it
maintains, within error bars, the value of the pure model, i.e.
$\nu=1$, in agreement with reference~\cite{fytas-08b}.
Turning now to the random bond square SAF model, the results
obtained here for the critical temperature, $T_{c}=1.978(8)$, and
correlation length exponent, $\nu=1.084(22)$, are also in good
agreement with our recent study of this model~\cite{fytas-08c},
where from the shift behavior of various pseudocritical
temperatures, the values $T_{c}=1.980(9)$ and $\nu=1.080(20)$ were
obtained.

In the corresponding insets of figure~\ref{fig:3} we illustrate
the scaling of the sample-to-sample fluctuations of the
sample-averaged pseudocritical temperatures of the magnetic
susceptibility, shown in the main panel of this figure, in a
log-log scale. Following Aharony and Harris~\cite{AH-96} (see also
the discussion in Section~\ref{sec:2}), we assume that the square
of these sample-to-sample fluctuations scales with the linear size
$L$ according to $L^{-n}$ we obtain, from the excellent fittings
for both models, the values $n=1.98(3)$ and $n=1.83(4)$,
respectively. For the case of the random bond SAF model the
situation is quite clear. The system is described by a new random
fixed point, so that $n=2/\nu$, and the estimate for $\nu$ is
$1.093(24)$, in agreement with the estimation $1.084(22)$ provided
above from the shift behavior of the corresponding pseudocritical
temperature and also with the traditional estimations presented in
reference~\cite{fytas-08c}. The result $n=1.98$ for the random
Ising model has within error bars the value of the space dimension
$d=2$ and thus provides further strong indications that, the
random bond version of the Ising model is governed by the pure
fixed point. An analogous study has been performed by Tomita and
Okabe~\cite{TO-01} for the 2d site diluted Ising model, where also
an exponent $n\approx 2$ has been estimated via the scaling of the
sample-to-sample fluctuations of the pseudocritical temperatures.
Thus, figure~\ref{fig:3} illustrates an important aspect of the
critical behavior of disordered systems that can be determined by
studying their self-averaging properties.

\section{Conclusions}
\label{sec:4}

In the present paper we have investigated the effects induced by
the presence of quenched bond randomness on the critical
self-averaging properties of two Ising spin models in 2d, namely
the regular Ising model and the case of competing interactions in
a generalized square Ising model. Our comparative study uncovered
the sensitivity of the microscopic competing interactions,
responsible for the SAF ordering, to bond randomness. This is
manifested in the much stronger violation of self-averaging of the
magnetic susceptibility and the large fluctuations of the
individual corresponding pseudoctritical temperatures. Moreover, a
detailed finite-size scaling analysis of the width of the
distribution of the sample-dependent pseudocritical temperatures
of the magnetic susceptibility verified the theoretical
expectations of Aharony and Harris and provided an alternative
approach of extracting the value for the correlation length
exponent. Thus, the generally undesirable feature of lack of
self-averaging in quenched random systems has been turned here
into a useful tool that contributes to our understanding of the
nature of phase transitions of these systems by providing an
alternative approach to criticality.

\ack{Research supported by the special Account for Research Grants
of the University of Athens under Grant No. 70/4/4071. N G Fytas
acknowledges financial support by the Alexander S. Onassis Public
Bene¯t Foundation.}

\end{document}